# Colossal Magnetoresistance in the $Mn^{2+}$ Oxypnictides $NdMnAsO_{1-x}F_x$

E. J. Wildman[1], J. M. S. Skakle[1], N. Emery[2] and A. C. Mclaughlin[1]*

[1]Department of Chemistry, University of Aberdeen, Meston Walk, Aberdeen AB24 3UE, UK

[2] Institut de Chimie et des Materiaux Paris Est, ICMPE/GESMAT, UMR 7182 CNRS-Universite Paris Est Creteil, CNRS 2 rue Henri Dunant, 94320 Thiais, France

**Abstract**

Colossal magnetoresistance (CMR) is a rare phenomenon in which the electronic resistivity of a material can be decreased by orders of magnitude upon application of a magnetic field. Such an effect could be the basis of the next generation of magnetic memory devices. Here we report CMR in the antiferromagnetic oxypnictide $NdMnAsO_{1-x}F_x$ as a result of competition between an antiferromagnetic insulating phase with strong electron correlations and a paramagnetic semiconductor upon application of a magnetic field. The discovery of CMR in antiferromagnetic $Mn^{2+}$ oxypnictide materials could open up an array of materials for further investigation and optimisation for technological applications.

**Introduction**

CMR has been observed in manganese oxides such as the perovskite $La_{1-x}Sr_xMnO_3$ ([1, 2, 3, 4]) and pyrochlore $Tl_2Mn_2O_7$ [5, 6]. Magnetoresistance is defined as the change in electrical resistivity, $\rho$, upon applying a magnetic field, H so that MR= $(\rho(H)-\rho(0))/\rho(0)$, where $\rho(0)$ and $\rho(H)$ are equal to the resistivity in zero and applied field respectively. Reductions in $\rho$ of up to 99.9% have been observed in thin films of $La_{0.67}Ca_{0.33}MnO_x$ in a 6 T magnetic field. The complete theory of CMR in manganese oxides is not yet established, but the largest magnetoresistance is evidenced in the vicinity of the paramagnetic to ferromagnetic transition, where a drop in electronic resistivity is also observed. $Mn^{3+}/Mn^{4+}$ mixed valency is present in manganese perovskites $A_{1-x}B_xMnO_3$ [1-4] (A is a trivalent element such as $La^{3+}$, $Nd^{3+}$ and B is a divalent element such as $Ca^{2+}$, $Sr^{2+}$) and it has been proposed that a dynamic Jahn-Teller effect [7], the double exchange (DE) mechanism and spatial electronic phase separation [8, 9] are all essential for the CMR mechanism. The magnetotransport properties of the pyrochlore manganite $Tl_2Mn_2O_7$ [5, 6] appear similar to the doped perovskite; however $Mn^{4+}$ is tetravalent in this material so that CMR cannot be related to the dynamic Jahn-Teller effect or the DE mechanism. Instead it has been proposed that the CMR arises from carrier localisation due to strong spin fluctuations [10]. CMR has also been reported in other ferromagnetic materials, for example $FeCr_2S_4$ [11] and $EuO$ [12] where the magnetotransport properties are very similar to the manganites. The recent discovery of high temperature superconductivity in pnictides such as $LnFeAsO_{1-x}F_x$ [13, 14] has led to great interest in these materials. Here we report an investigation of the $NdMnAsO_{1-x}F_x$ analogue which shows antiferromagnetic order of the $Mn^{2+}$ and $Nd^{3+}$ spins below 356 K and 23 K respectively. Colossal magnetoresistance is observed at low temperature and we show that this is as a result of a second order phase transition in field from an antiferromagnet (AF) to a paramagnet (P), so that the CMR arises due to competition between AF and P phases with very different electron transport.

**Results**

Rietveld refinement of 290 K and 4 K synchrotron X-ray powder diffraction data (Supplementary Fig. 1 and Table 1) recorded for $NdMnAsO_{0.95}F_{0.05}$ confirms the ZrCuSiAs crystal structure previously reported for the parent compound at both temperatures [15, 16, 17] which consists of a tetrahedral $[MnAs]^-$ layer sandwiched between slabs of insulating $[NdO/F]^+$ (space group *P4/nmm*; $a$ = 4.048697(3) Å and $c$ = 8.89654(1) Å at 290 K) (Fig. 1a). Figure 1b shows the variation of magnetic susceptibility with temperature for $NdMnAsO_{1-x}F_x$ ($x$ = 0.05). Two magnetic transitions are observed below ~ 23 K. There is no divergence of zero field cooled (ZFC) and field cooled (FC) susceptibility at either of these transitions. The inset to Figure 1b shows the field variation of the magnetisation, M, in which there is no evidence of a ferromagnetic component upon increasing the magnetic field, $\mu_0H$, from 0 – 7 T but there is a subtle change in slope at 3.5 T.

Neutron diffraction data shows that below 356 (2) K, antiferromagnetic order of the $Mn^{2+}$ moments is observed with the same magnetic structure as previously reported for the parent compound NdMnAsO [15, 16, 17] so there is no appreciable change in $T_N$ (Mn) with 5% F doping [17]. The temperature variation of the $Mn^{2+}$ moment is displayed in Fig. 1c; this transition is not evidenced in the magnetic susceptibility data (Fig. 1b). Below 23 K ($T_{N1}$ (Nd)), antiferromagnetic order of the $Nd^{3+}$ moments is evidenced with spins aligned parallel to the basal plane (Fig. 1c (inset), Supplementary Fig. 2). At the same time a spin reorientation of the $Mn^{2+}$ moments into the *ab* plane is observed as previously reported for NdMnAsO [16, 17] so that at 20 K, $T_{SR}$, the $Mn^{2+}$ moments are fully aligned parallel to the basal plane. At 13 K, a subtle second transition of the $Nd^{3+}$ spins ($T_{N2}$ (Nd)) is observed (Fig. 1c) which corresponds to antiferromagnetic ordering of larger moments of the $Nd^{3+}$ spins with the same propagation vector. Similar magnetic ordering of $Nd^{3+}$ has been observed in NdCoAsO [18]. There is no evidence of magnetic phase separation upon cooling as previously observed in CMR manganites [19].

Figure 2a shows the temperature variation of the 7 T magnetoresistance of $NdMnAsO_{1-x}F_x$ for x = 0.050, 0.065 and 0.080. For all samples –MR is observed below ~75 K which decreases exponentially as the temperature is reduced. Below $T_{SR}$ the magnetoresistance drops sharply and decreases further as the temperature is lowered. The temperature variation of the resistivity in H = 0 T and H = 7 T is displayed in Supplementary Figure 3. For all samples the resistivity can be modelled by three-dimensional variable range hopping (VRH) of the carriers below 75 K [20] (phonon assisted tunnelling of electrons between localised state so that the resistivity, ρ, is defined as $\rho=\rho_0 exp(T_0/T)^{0.25}$). The inset to Fig. 2a evidences a subtle electronic transition at $T_{SR}$ as the resistivity data can no longer be fit well to the three-dimensional VRH equation below 20 K. This can be modelled as a crossover from three-dimensional to one dimensional Efros Shklovskii VRH [21]; an excellent fit is obtained to this model down to the lowest temperature measured.

Figure 2b shows the field variation of the –MR for NdMnAsO$_{0.95}$F$_{0.05}$ at several different fixed temperatures. Below T$_{SR}$ the –MR can be fit to the equation –MR = A + BH + CH$^2$ (A, B and C are constants) between 0.5 – 3.5 T (Supplementary Table 2) but deviates from this equation at higher fields. Above T$_{SR}$, the data can be fit to the same equation. The -MR is reversible with field and reaches a value of -95 % at 3 K in a 9 T field which is comparable to perovskite and pyrochlore manganites [1-10]. This is surprising given the very different electronic and magnetic properties of NdMnAsO$_{0.95}$F$_{0.05}$ compared to the perovskite and pyrochlore manganites and suggests a novel CMR mechanism. The MR observed here cannot be attributed to spin polarised tunnelling across domain or phase boundaries, where typically an appreciable low field (< 1 T) MR is detected [22, 23]. Furthermore, there is no correlation between the field variation of the magnetisation and the MR. Figure 3a shows a portion of the 4 K neutron diffraction pattern where upon applying a magnetic field it is apparent that the intensity of the [100] magnetic peak decreases. There is no change in intensity of any of the other magnetic or nuclear peaks and hence there is no evidence of an increasing ferromagnetic component with field (Supplementary Fig. 4). Consequently there is no canting or reorientation of the spins with field, which is in agreement with the variable field magnetic susceptibility data (Fig 1b). The field variation of the magnetic structure was determined by Rietveld refinement of the neutron diffraction data, so that the decreased intensity of the [100] magnetic peak can be modelled by a reduction in both the staggered Nd$^{3+}$ and Mn$^{2+}$ moments with increasing field. The Nd$^{3+}$ and Mn$^{2+}$ moments are reduced from 2.09 (3) $\mu_B$ to 1.36 (3) $\mu_B$ and from 3.83 (2) $\mu_B$ – 3.56 (2) $\mu_B$ respectively upon increasing $\mu_0$H from 0 – 5 T. Upon returning the field to 0 T, the Mn$^{2+}$ and Nd$^{3+}$ moments refine to 3.86 (2) $\mu_B$ and 2.10 (3) $\mu_B$ respectively so that the reduction in magnetic moments with field is reversible.

**Discussion**

NdMnAsO is non-stoichiometric and has a Nd occupancy of 0.970(2), but the holes generated by this non-stoichiometry are trapped by the cation disorder (Anderson localisation) [16]. Rietveld refinement of synchrotron X-ray powder diffraction data demonstrates that the stoichiometry of the 5% doped compound is $Nd_{0.97}MnAsO_{0.95}F_{0.05}$. NdMnAsO is a semiconductor and exhibits three dimensional variable range hopping of the carriers between 4 – 380 K. Below 200 K a sizeable –MR has been reported equal to -12 % in a 5 T field as a result of a reduction in quantum destructive interference with field [15, 16]. Upon substitution of $O^{2-}$ with $F^-$, electrons are doped into the system and it is assumed that the conduction carriers are confined to the covalent $[MnAs]^-$ layer as opposed to the insulating $[LaO/F]^+$ slab. The electronic properties are observed to change remarkably with 5% $F^-$ substitution and multiple electronic transitions are evidenced. $NdMnAsO_{0.95}F_{0.05}$ is a degenerate semiconductor, so that at high temperature metal-like temperature dependence of the resistivity is observed, but between 160 K and 80 K the electronic behaviour is dominated by thermally activated charge carriers across a band gap so that $\rho = \rho_0 \exp(E_g/2kT)$ ($E_g$ = 23 meV ). Upon cooling to 75 K three dimensional VRH hopping of the carriers is observed, but at $T_{SR}$ a crossover to Efros Shklovskii VRH ($\rho=\rho_0\exp(T_{ES}/T)^{1/2}$) is detected. This suggests that reorientation of $Mn^{2+}$ spins into the basal plane at 20 K results in enhanced Coulomb correlations between localised electrons so that a soft Coulomb gap pinned at the Fermi level opens up and Efros Shklovskii (ES) VRH is observed (Fig. 2a), i.e. that the spin orientation has a strong influence on the hopping mechanism so that when the spins align in the basal plane, the electron correlations are enhanced and the transport is diminished. Upon application of a magnetic field both the $Mn^{2+}$ and $Nd^{3+}$ ordered moments reduce (Fig. 3) in magnitude, the electron correlations diminish and an increase in electronic transport is evidenced so that a large drop in magnetoresistance is observed at $T_{SR}$ (Fig. 2a). Further inspection of the resistivity data recorded in a 7 T field shows that ES VRH is no longer observed below 20 K. Instead two-dimensional VRH of the carriers is detected between 18 – 3 K ($\rho=\rho_0\exp(T_0/T)^{1/3}$) (Supplementary Fig. 5).

Figure 3b shows that the reduction of the staggered $Nd^{3+}$ moment with field can be well fit by the standard expression $M^2 = M_0^2[1- (H/H_m)^\gamma]$ [24] ($M_0$ is the moment in zero field, $H_m$ is the critical field for the magnetic order and $\gamma = 2$), which describes the magnetic field dependence of the magnetic moment for a second order phase transition to the paramagnetic phase and is derived from Ginzburg-Landau theory. Fits to this equation result in $H_m$ = 6.3(5) T so that antiferromagnetic order of the $Nd^{3+}$ spins is expected to vanish at this field. A slightly different variation of the staggered $Mn^{2+}$ moment is evidenced (Fig. 3b (inset)) so that the data between $\mu_0H$ = 0 – 3.5 T can be fit to the same equation with $H_m$ =11.1(2) T. However at 3.5 T there is a clear inflection point. This subtle change at 3.5 T is also evidenced in the field evolution of the magnetisation (Fig. 1b) so that the uniform magnetisation incorporates the field alignment of paramagnetic $Nd^{3+}$ and $Mn^{2+}$ spins which are evidenced upon application of a magnetic field. This is clearly not saturated up to 7 T. The 3.5 T anomaly is also detected in the magnetoresistance data (Fig. 2b) and it is not possible to fit the field variation of the MR or the $Mn^{2+}$ moment by a single function over the field range measured.

Figure 4 shows that there is in fact a correlation between the field variation of the staggered $Mn^{2+}$ moment and the observed magnetoresistance across the field range measured (0 – 5 T) at 4 K; where the MR is related to the $Mn^{2+}$ moment so that $-MR = \left(\frac{\Delta M}{C}\right)^{1/2}$ where $\Delta M$ = M(0) – M(H) where M(H) is the $Mn^{2+}$ moment in field, M(0) is the $Mn^{2+}$ moment when $\mu_0H$ = 0 T and C is a constant (0.4 $\mu_B$ at 4 K) which equates to the moment reduction that would theoretically result in a -MR of 100%. The same relation is not detected with the field reduction of the $Nd^{3+}$ moment. This clearly shows that it is the field-suppression of the basal plane staggered $Mn^{2+}$ moment which is the origin of the CMR observed in $NdMnAsO_{1-x}F_x$ below $T_{SR}$ so that the electron correlations are weaker in the paramagnetic phase, resulting in dramatically increased electron transport. It is remarkable that only a 7% reduction of the $Mn^{2+}$ moment results in an 86% reduction in the electronic resistivity. Presumably the MR observed above $T_{SR}$ has the same origin, but the difference in the electronic transport of the antiferromagnetic and paramagnetic phases is less significant when the $Mn^{2+}$ spins are aligned parallel to *c*. Large negative magnetoresistances have previously been reported for the antiferromagnetic Zintl compound $Eu_{14}MnBi_{11}$ as a result of increased ferromagnetic fluctuations with field [25]; at all temperatures below $T_N$ there is a linear correlation between the –MR and the raw moment (which increases with field). The same correlation is not observed for $NdMnAsO_{1-x}F_x$ in which there is no evidence of ferromagnetic fluctuations.

The origin of the second order phase transition, which results in the reduction of both the $Mn^{2+}$ and $Nd^{3+}$ moments with increasing magnetic field, is as yet unknown. In most antiferromagnetic materials, the exchange coupling is stronger than other internal interactions so that strong magnetic fields are required to destroy the antiferromagnetic order and align the sublattice magnetisations with the field (spin flip transition into the field-aligned paramagnetic state). The application of the smaller magnetic fields applied here should result in only a slight distortion from the antiparallel alignment. Uniaxial antiferromagnets such as $NiCl_2.6H_2O$[26] and $CePb_3$[27] are well documented to display spin-flop transitions or spin rotations with field, but there is no evidence of such transitions from the variable field susceptibility and neutron diffraction data. The $Mn^{2+}$ spins order antiferromagnetically below 356 K in $NdMnAsO_{0.95}F_{0.05}$ which suggests that there is strong antiferromagnetic exchange coupling between $Mn^{2+}$ moments so that the second order phase transition to a paramagnetic phase with field is unexpected and suggests the presence of an antiferromagnetic instability.

It is possible that the competing single-ion anisotropy which results in the low temperature spin reorientation transition [16, 17] is also the origin of the field induced paramagnetism. An alternative explanation is that there is a hidden order parameter, indeed the decrease of magnetic intensity of the $Mn^{2+}$ moment with field does not follow a simple power law, but shows a clear inflection point at 3.5 T (Fig. 3b). Such behaviour has previously been observed in the heavy fermion material URu$_2$Si$_2$ as a result of a linear coupling between a hidden order parameter and the antiferromagnetic moment [28], and it appears that a similar complex underlying physics is also present in NdMnAsO$_{1-x}$F$_x$.

In conclusion we have shown a novel mechanism of CMR in the oxypnictide material NdMnAsO$_{1-x}$F$_x$. CMR in manganites arises when a magnetic field enhances ferromagnetic alignment of spins at the paramagnetic – ferromagnetic (disorder – order) boundary. Here we show that CMR is also possible in Mn$^{2+}$ pnictides at the antiferromagnetic – paramagnetic (order-disorder) transition in field. These results demonstrate that it is possible to observe CMR solely as a result of competition between different phases in field which have divergent electronic properties. CMR is not observed in the parent compound, which suggests that the spin polarised charge carriers are the electrons generated by the substitution of F$^-$ for O$^{2-}$. Finally we note that upon replacement of Mn$^{2+}$ with Fe$^{2+}$ in NdMnAsO$_{1-x}$F$_x$ high temperature superconductivity (HTSC) is established. The discovery of HTSC in pnictides is remarkable and the report of CMR in the Mn$^{2+}$ analogue further highlights the exotic physics of transition metal pnictides.

**Experimental**

Polycrystalline samples of Nd$_{0.97}$O$_{0.95}$MnAsF$_{0.05}$ were prepared *via* a two step solid-state reaction. Pieces of rare earth Nd (Aldrich 99.9%) and As (Alfa Aesar 99.999%) were reacted at 900°C in a quartz tube sealed under vacuum. The precursor was then reacted with stoichiometric amounts of MnO$_2$, Mn and MnF$_2$ (Aldrich 99.99%). The powders were ground and pressed into pellets of 10 mm diameter and placed in a Ta crucible. The pellets were then sintered at 1150°C for 48 hours in an evacuated quartz tube.

High resolution synchrotron X-ray powder diffraction patterns were recorded on the ID31 beamline at the ESRF, Grenoble, France at 290 K with a wavelength of 0.3999 Å. The powder sample was inserted into a 0.5mm diameter borosilicate glass capillary and spun at ~1Hz. The patterns were collected between 2° < 2$\vartheta$ < 53°.

Powder neutron diffraction data were recorded using the D20 high intensity diffractometer at the Institute Laue Langevin (ILL, Grenoble, France). Neutrons of wavelength 2.4188 Å were incident on an 8 mm vanadium can contained in a vertical cryomagnet. Zero field data were recorded between 4-370 K and variable field data were recorded at 4 K between 0 – 5 T and again at 0 T with a collection time of 20 minutes per field/temperature. The temperature and field dependency of the magnetic structure of the compound was obtained by Rietveld refinement of the neutron data.

The temperature and field dependence of the electrical resistance were recorded using a Quantum Design physical property measurement system (PPMS) between 4 and 400 K and in magnetic fields up to 9 T.

The magnetic susceptibility was measured with a Quantum Design superconducting quantum interference device magnetometer (SQUID). Zero field cooled (ZFC) and field cooled (FC) measurements were recorded between 2-400 K in a field of 1000 Oe. Variable field data were recorded at 2 K between $\pm 7$ T.

**Acknowledgements** We thank M. Brunelli and A. Hill for assistance with the neutron and synchrotron experiments. We also acknowledge UK EPSRC (grant EP/F035225/1) and the Royal Society for financial support and STFC for beam time provision.

Correspondence and requests for materials should be addressed to Abbie Mclaughlin (a.c.mclaughlin@abdn.ac.uk).

**Figure captions**

**Figure 1. The crystal structure and magnetic properties of NdMnAsO$_{1-x}$F$_x$. a,** The crystal structure of NdMnAsO$_{0.95}$F$_{0.05}$ (Mn = green atoms, As = turquoise atoms, Nd = purple atoms, O and F = red atoms). **b,** Zero field cooled (ZFC) and field cooled (FC) variable temperature susceptibility data recorded for NdMnAsO$_{0.95}$F$_{0.05}$. The inset shows the field variation of the magnetisation; a change of slope is observed above 3.5 T as indicated by the arrow. **c,** The thermal variation of the Mn$^{2+}$ magnetic moment refined from neutron data with T$_N$ (Mn) = 356 K. The inset shows the variation of the Nd$^{3+}$ moment with temperature evidencing two magnetic transitions at 23 K and 13 K.

**Figure 2. Magnetotransport data for NdMnAsO$_{1-x}$F$_x$. a,** The variation of magnetoresistance with temperature for x = 0.050, 0.065 and 0.080 evidencing two magnetoresistive transitions at T$_{SR}$ and T$_{N2}$(Nd) for several NdMnAsO$_{1-x}$F$_x$ samples . The inset shows a crossover from three dimensional variable range hopping to one dimensional variable range hopping below T$_{SR}$; T$_{ES}$ = 922 K and T$_0$ = 4 x 10$^5$ K in the low and high temperature regions respectively. **b,** The field dependence of the magnetoresistance at selected temperatures between 3 – 30 K. The solid lines show fits to the equation –MR = A + BH + CH$^2$.

**Figure 3. Changes in magnetic structure with field. a,** A portion of the 4 K neutron diffraction pattern showing a reduction in intensity of the (100) magnetic diffraction peak with increasing magnetic field ($\mu_0$H) from 0 – 5 T. This corresponds to a reduction in both Mn$^{2+}$ and Nd$^{3+}$ moments with increasing $\mu_0$H. **b,** Variation of the normalised Nd$^{3+}$ moment with $\mu_0$H; the data is fit to the equation M$^2$ =M$_0^2$[1- (H/H$_m$)$^2$]. The inset shows the variation of the normalised Mn$^{2+}$ moment with $\mu_0$H; the data is fit to the equation M$^2$ =M$_0^2$[1- (H/H$_m$)$^2$] between 0 – 3.5 T and a clear inflection point is observed at 3.5 T.

**Figure 4. Variation of MR$^2$ with Mn$^{2+}$ moment.** The results show an empirical correlation between the observed −MR and the field reduced Mn$^{2+}$ moment recorded at 4 K with H = 0 − 5 T (every 0.5 T).

Fig. 1

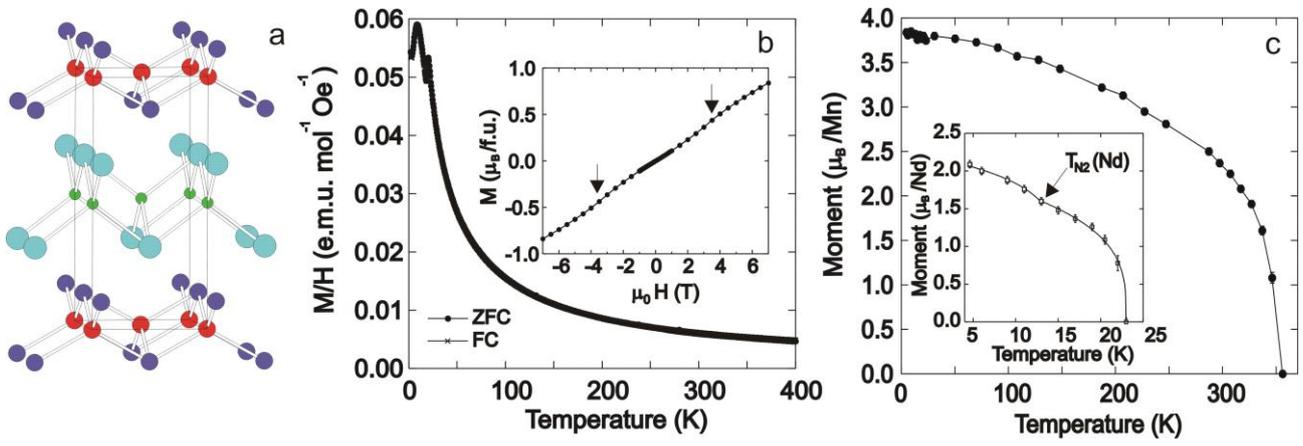

Fig. 2

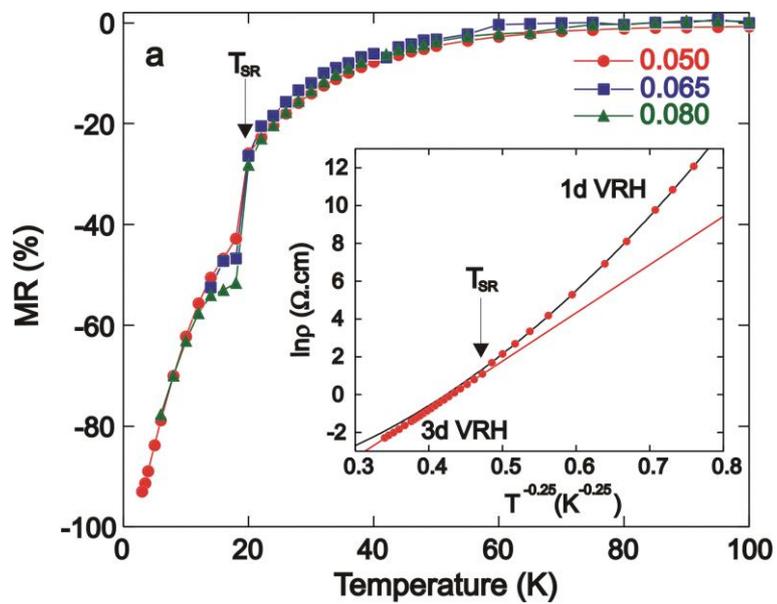

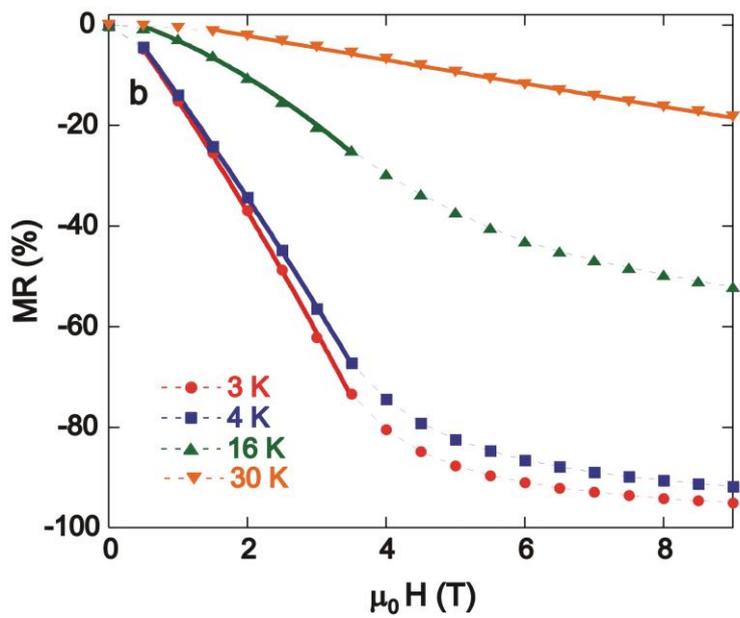

Fig. 3

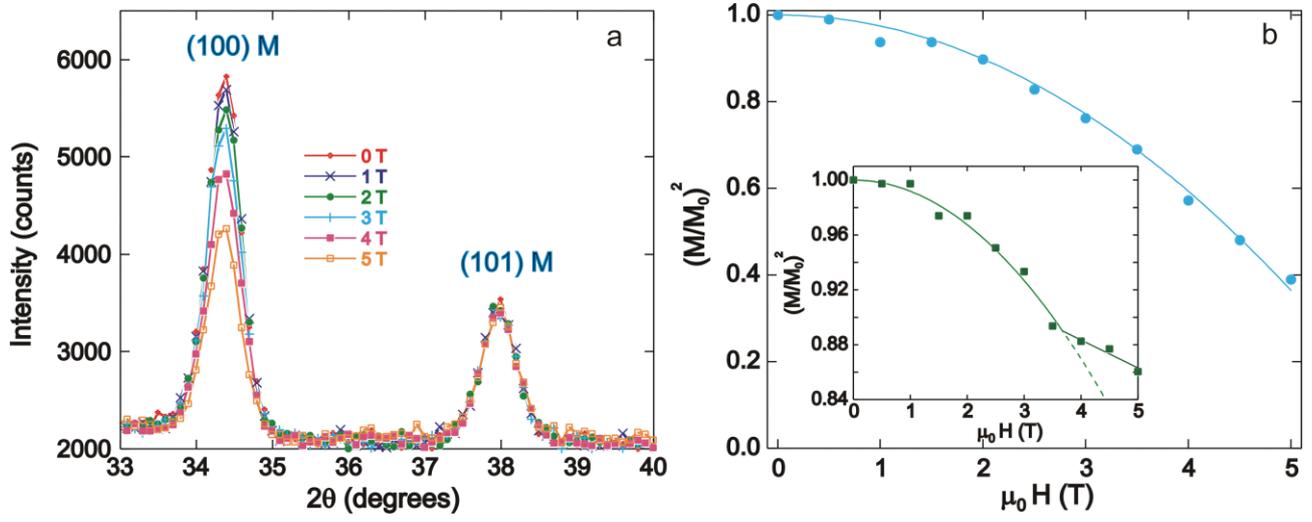

Fig. 4

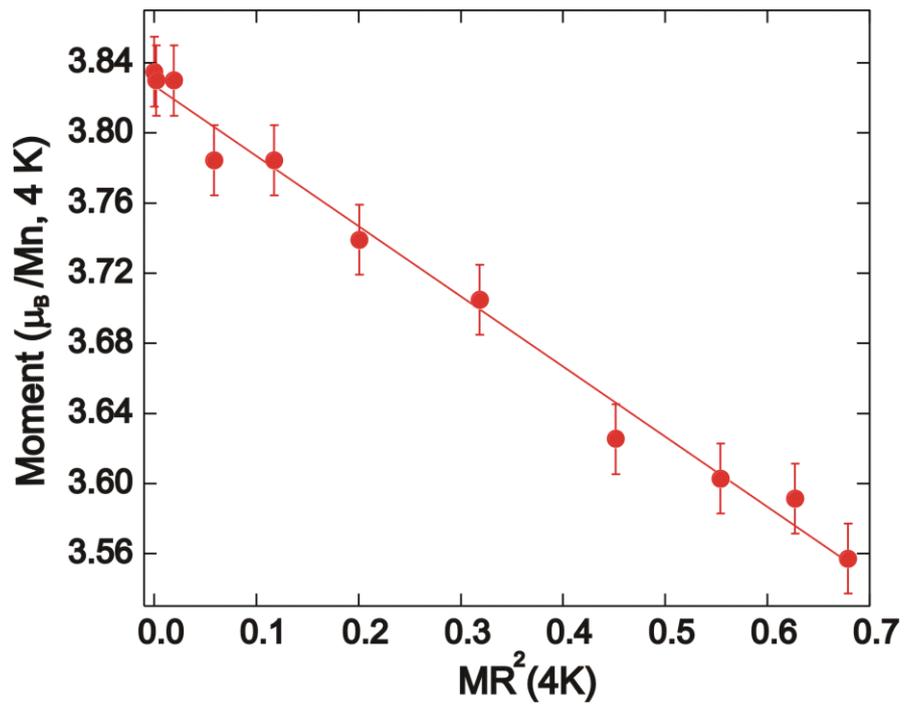